\documentclass{article}
\usepackage{amssymb}
\newtheorem{theorem}{Theorem}
\newtheorem{lemma}[theorem]{Lemma}
\newtheorem{corollary}[theorem]{Corollary}

\newcommand{\diff}[3]{\frac{\partial^{#3}#1}{\partial #2^{#3}}}

\def\d{\displaystyle}

\def\loc{{\hbox{\tiny loc}}}

\def\lap{\Delta}
\def\flow{\varphi}
\def\eps{\varepsilon}

\def\R{\mathbb{R}}
\def\N{\mathbb{N}}
\begin{document}
%-----------------------------------------------------------------------
\noindent
\textbf{\Large Generalized Flows, Intrinsic Stochasticity,\\ and
  Turbulent Transport}

~

\noindent 
\textsc{Weinan E and Eric Vanden Eijnden}

\noindent 
{\scriptsize Courant Institute of Mathematical Sciences, \\ New York
  University, New York, NY 10012}

\vskip3pt

\noindent
{\bf ABSTRACT~~~ The study of passive scalar transport in a turbulent
  velocity field leads naturally to the notion of generalized flows
  which are families of probability distributions on the space of
  solutions to the associated ODEs which no longer satisfy the
  uniqueness theorem for ODEs.  Two most natural regularizations of
  this problem, namely the regularization via adding small molecular
  diffusion and the regularization via smoothing out the velocity
  field are considered. White-in-time random velocity fields are used
  as an example to examine the variety of phenomena that take place
  when the velocity field is not spatially regular.  Three different
  regimes characterized by their degrees of compressibility are
  isolated in the parameter space.  In the regime of intermediate
  compressibility, the two different regularizations give rise to two
  different scaling behavior for the structure functions of the
  passive scalar.  Physically this means that the scaling depends on
  Prandtl number.  In the other two regimes the two different
  regularizations give rise to the same generalized flows even though
  the sense of convergence can be very different. The ``one force, one
  solution'' principle is established for the scalar field in the
  weakly compressible regime, and for the difference of the scalar in
  the strongly compressible regime which is the regime of inverse
  cascade.  Existence and uniqueness of an invariant measure is also
  proved in these regimes when the transport equation is suitably
  forced.  Finally incomplete self-similarity in the sense of
  Barenblatt-Chorin is established.  }

\vskip3pt
\noindent
\hrulefill
\vskip5pt

\section*{Introduction}

Recent efforts to understand the fundamental physics of hydrodynamic
turbulence have concentrated on the explanation of the observed
violations of Kolmogorov's scaling. These violations reflect the
occurrence of large fluctuations in the velocity field on the small
scales, a phenomenon referred to as intermittency.  Some progress in
the understanding of intermittency has been achieved recently through
the study of simple model problems that include Burgers equation
\cite{ekhma97,eva99} and the passive advection of a scalar by a
velocity field of known statistics
\cite{shsi93,chfakole95,gaku95,bale98}.  This
paper is a summary of the many interesting mathematical issues that
arise in the problem of passive scalar advection together with our
understanding of these issues. We put some of our results in the
perspective of a new phenomenological model proposed recently by
Barenblatt and Chorin \cite{bach97,bach98} using the formalism
of incomplete self-similarity.

\section*{Generalized Flows}

Consider the transport equation for the scalar field
$\theta^\kappa({x},t)$ in $\R^d$:
\begin{equation} 
  \label{eq:transp}
  \diff{\theta^\kappa}{t}{} +({u}({x},t) \cdot \nabla) \theta^\kappa 
  = \kappa \lap \theta^\kappa.
\end{equation}
We will be interested in $\theta^\kappa$ in the limit as $\kappa\to0$.
It is known from classical results that if ${u}$ is Lipschitz continuous
in ${x}$, then as $\kappa\to0$, $\theta^\kappa$ converges to $\theta$,
the solution of
\begin{equation} 
  \label{eq:scallimitLip1}
  \diff{\theta}{t}{} + ({u}({x},t) \cdot \nabla) \theta=0.
\end{equation}
 Furthermore, if we define $\{\flow_{s,t}({x})\}$ as  the flow
generated by the velocity field ${u}$, satisfying the ordinary
differential equations (ODEs)
\begin{equation} 
  \label{eq:deter}
  \frac{d \flow_{s,t}({x})}{dt} = {u}(\flow_{s,t}({x}),t), 
  \qquad \flow_{s,s}({x}) = {x},
\end{equation}
for $s<t$, then the solution of the transport equation in
\ref{eq:scallimitLip1} for the initial condition
$\theta^\kappa({x},0) = \theta_0({x})$ is given by
\begin{equation}
  \label{eq:limkapp1}
  \theta({x},t)=\theta_0(\flow^{-1}_{0,t} ({x}))
  = \theta_0(\flow_{t,0} ({x})).
\end{equation}
This classical scenario breaks down when ${u}$ fails to be Lipschitz
continuous in ${x}$, which is precisely the case for fully developed
turbulent velocity fields. In this case Kolmogorov's theory of
turbulent flows suggests that ${u}$ is only H\"older continuous with
an exponent roughly equal to $\frac{1}{3}$ for $d=3$.  In such
situations the solution of the ODEs in \ref{eq:deter} may fail to be
unique \cite{rem1}, and we then have to consider probability
distributions on the set of solutions in order to solve the transport
equation in \ref{eq:scallimitLip1}. This is the essence of the notion
of generalized flows proposed by Brenier \cite{bre89,bre93}
(see also \cite{dima86,shn94}).

There are two ways to think about probability distributions on the
solutions of the ODEs in \ref{eq:deter}. We can either think of it as
probability measures on the path-space (functions of~$t$) supported by
paths which are solutions of \ref{eq:deter}, or we can think of it as
transition probability at time $t$ if the starting position at
time~$s$ is~${x}$.  In the classical situation when ${u}$ is Lipschitz
continuous, this transition probability degenerates to a point mass
centered at the unique solution of \ref{eq:deter}. When Lipschitz
condition fails, this transition probability may be non-degenerate and
the system in \ref{eq:deter} is intrinsically stochastic.

There is a parallel story for the case when~${u}$ is a
white-in-time random process 
defined on a probability space $(\Omega,\mathcal{F},\mathcal{P})$.
We will denote the elements in $\Omega$ by $\omega$ and  indicate the
dependence on realization of the random velocity field by a super- or
a subscript~$\omega$.  In connection with the transport equation in
\ref{eq:scallimitLip1}, it is most natural to consider the stochastic
ODEs
\begin{equation}
  \label{eq:ODF}
  d\flow^{\omega}_{s,t}({x})
  ={u}(\flow^{\omega}_{s,t}({x}),t)dt,
  \qquad  \flow^{\omega}_{s,s}({x}) = {x},
\end{equation}
in Stratonovich sense. In this case, it is shown \cite{kun90} that
if the local characteristic of ${u}$ is spatially twice
continuously differentiable, then the system in \ref{eq:ODF} has a
unique solution.  Such conditions are not satisfied by typical
turbulent velocity fields on the scale of interest. 
When the regularity condition on ${u}$ fails, there are at least
two natural ways to regularize \ref{eq:deter} or \ref{eq:ODF}. The
first is to add diffusion:
\begin{equation}
  \label{eq:kapreg}
  d\flow^{\omega,\kappa}_{s,t}({x})
  ={u}(\flow^{\omega,\kappa}_{s,t}({x}),t) dt
  +\sqrt{2\kappa} d{\beta}(t),
\end{equation}
and consider the limit as $\kappa\to0$. We will call this the
$\kappa$-limit.  The second is to smooth out the velocity field. Let
$\psi_\eps$ be defined as $\psi_\eps({x})=\eps^{-d}\psi(x/\eps),$
where $\psi$ is a standard mollifier: $\psi\geq0$, $\int_{\R^d}\psi
d{x}=1$, $\psi$ decays fast at infinity.  Let ${u}^{\eps}={u}\star
\psi_\eps$ and consider
\begin{equation}
  \label{eq:epsreg}
  d\flow^{\omega,\eps}_{s,t}({x})=
  {u}^{\eps}(\flow^{\omega,\eps}_{s,t}({x}),t) dt,
\end{equation}
in the limit as $\eps\to0$. We will call this the $\eps$-limit.
Physically $\kappa$ plays the role of molecular diffusivity, $\eps$
can be thought of as a crude model of the viscous cut-off scale.
The $\kappa$-limit corresponds to the situation when the
Prandtl number, defined here as the ratio of $\eps$ and $\kappa$,
 tends to zero, $Pr\to0$, whereas the $\eps$-limit
corresponds to the situation when the Prandtl number diverges,
$Pr\to\infty$. The following questions naturally arise:
\begin{itemize}
\item[(Q1)] How do the flows and the passive scalar behave
statistically in the $\kappa$- and $\eps$-limits?
\item[(Q2)] Does there exist a unique statistical steady state when the
  transport equation in \ref{eq:transp} is suitably forced?
\item[(Q3)] What are the statistical and geometrical properties of solutions
in the statistical steady state?
\end{itemize}
Below we address these questions  on a specific model introduced
by Kraichnan \cite{kra68}. 
 
Before proceeding further, we relate the regularized flows in
\ref{eq:kapreg}, \ref{eq:epsreg} to the solutions of the transport
equations.  Consider the $\kappa$-regularization first. It is
convenient to introduce the backward transition probability 
\begin{equation}
  \label{eq:gkappa}
  g_\omega^\kappa ({x},t|d{y},s) = \mathbf{E}_{\beta}
  \delta({y} -\flow^{ \omega,\kappa}_{t,s}({x}))d{y}, 
  \quad s<t,
\end{equation}
where the expectation is taken with respect to ${\beta}(t)$, and
$\flow^{ \omega,\kappa}_{t,s}({x})$ is the flow inverse to
$\flow^{ \omega,\kappa}_{s,t}({x})$ defined in \ref{eq:kapreg}
(i.e.  $\flow^{ \omega,\kappa}_{s,t}({x})$ is the forward flow
and $\flow^{ \omega,\kappa}_{t,s}({x})$ is the backward flow).
The action of $g_\omega^\kappa$ generates a semi-group of
transformation
\begin{equation}
  \label{eq:Skappa}
  S^{\omega,\kappa}_{t,s} \psi({x})  = 
  \int_{\R^d}  \psi({y}) g_\omega^\kappa ({x},t|d{y},s),
\end{equation}
for all test functions~$\psi$.  $\theta_\omega^\kappa({x}, t) =
S^{\omega,\kappa}_{t,s} \psi({x})$ solve the transport equation in
\ref{eq:transp} for the initial condition
$\theta^\kappa_\omega({x}, s) = \psi({x})$.  Similarly, for
the flow in \ref{eq:epsreg}, define
\begin{equation}
  \label{eq:thetageps}
  S_{t,s}^{\omega,\eps}\psi({x}) = 
  \psi(\flow^{\omega,\eps}_{t,s}({x})), \qquad s<t.
\end{equation}
$\theta_\omega^\eps({x}, t)= S_{t,s}^{\omega,\eps}\psi({x})$
solves the transport equation
\begin{equation}
  \label{eq:transpeps}
    \diff{\theta^\eps}{t}{} + 
    ({u}^{\eps}({x},t) \cdot \nabla)\theta^\eps=0,
\end{equation}
with initial condition $\theta({x}, s) = \psi({x})$.  Similar
definitions can be given for forward flows but we will restrict
attention to the backward ones since we are primarily interested in
scalar transport. The results given below generalize trivially to
forward flows.

\section*{Kraichnan Model}

In \cite{kra68} Kraichnan introduced one of the simplest model of
passive scalar by considering the advection by a Gaussian, spatially
non-smooth and white-in-time velocity field. The fact that
white-in-time velocity fields may exhibit intermittency was first
recognized by Majda \cite{maj93,makr99}. Definitive work on
Kraichnan model has been done afterwards in
\cite{shsi93,chfakole95,gaku95,bale98}.

We will consider a generalization of Kraichnan model introduced in
\cite{gave00} (see also \cite{lera99}). The velocity ${u}$ is
assumed to be a statistically homogeneous, isotropic and stationary
Gaussian field with mean zero and covariance
\begin{equation}
  \label{eq:covdef}
  \mathbf{E}\, u_\alpha({x},t)u_\beta({y},s)=
  (C_0 \delta_{\alpha\beta}-c_{\alpha\beta}({x}-{y}))\delta(t-s).
\end{equation}
We assume that ${u}$ has a correlation length~$\ell_0$, i.e. the
covariance in \ref{eq:covdef} decays fast for
$|{x}-{y}|>\ell_0$. Consequently $c_{\alpha\beta}({x})\to
C_0 \delta_{\alpha\beta}$ as $|{x}|/\ell_0\to\infty$.  On the
other hand, we will be mainly interested in small scale phenomena for
which $|{x}|\ll\ell_0$.  In this range, we take
$c_{\alpha\beta}({x}) = d_{\alpha\beta}({x}) +
O(|{x}|^2/\ell_0^2)$ with
\begin{equation}
  \label{eq:ddef}
  d_{\alpha\beta}({x})=
  Ad^P_{\alpha\beta}({x})+
  Bd^S_{\alpha\beta}({x}),
\end{equation}
and
\begin{equation}
  \label{eq:DpDs}
  \begin{array}{l}
    {\d d^P_{\alpha\beta}({x})=D
      \left(\delta_{\alpha\beta}
        +\xi \frac{x_\alpha x_\beta}{|{x}|^2}\right)|{x}|^{\xi},}\\
    {\d d^S_{\alpha\beta}({x})=D
      \left((d+\xi-1)\delta_{\alpha\beta}
        -\xi \frac{x_\alpha x_\beta}{|{x}|^2}\right)|{x}|^{\xi}.}
  \end{array}
\end{equation}
$D$ is a parameter with dimension
$[\hbox{length}]^{2-\xi}[\hbox{time}]^{-1}$.  The dimensionless parameters
$A$ and $B$ measure the divergence and rotation of the field
${u}$. $A=0$ corresponds to incompressible fields with
$\nabla\cdot {u}=0$. $B=0$ corresponds to irrotational fields with
$\nabla \times {u}=0$. The parameter $\xi$ measures the spatial
regularity of ${u}$. For $\xi\in (0,2)$, the local characteristic of
${u}$ fails to be twice differentiable and this fact has important
consequences on both the transport equation in \ref{eq:scallimitLip1}
and the systems of ODEs in \ref{eq:deter} or \ref{eq:ODF}.

Existing physics literature concentrates on the $\kappa$-limit for
Kraichnan model. Let $\mathcal{S}^2=A+(d-1)B$, $\mathcal{C}^2=A$,
$\mathcal{P}=\mathcal{C}^2/\mathcal{S}^2$.  $\mathcal{P}\in[0,1]$ is a
measure of the degree of compressibility of ${u}$. The pioneering
work of Gaw\c edzki and Vergassola \cite{gave00} (see also
\cite{lera99}) identifies two different regimes for the
$\kappa$-limit:

1.  The strongly compressible regime when $\mathcal{P}\geq d/\xi^2$.
In this regime $g^\kappa_\omega$ converges to a flow of maps, i.e.
there exists a two-parameter family of maps
$\{\flow^{\omega}_{t,s}({x})\}$ such that
\begin{equation}
  \label{eq:twoparam}
  g^\kappa_\omega({x},t|d{y},s)\to
  \delta({y}-\flow^{\omega}_{t,s}({x})) d{y}.
\end{equation}
Moreover particles have finite probability to coalesce under the flow
of $\{\flow^{\omega}_{t,s}({x})\}$.
In other words the flow is not invertible.

2. When $\mathcal{P} < d/\xi^2$, $g^\kappa_\omega$ converges
to a ``generalized stochastic flow''
\begin{equation}
  \label{eq:genstoch}
  g^\kappa_\omega({x},t|d{y},s)\to 
  g_\omega({x},t|d{y},s),
\end{equation}
and the limit $g_\omega$ is a nontrivial probability distribution in
${y}$.  This means that the image of a particle under the flow
defined by the velocity field ${u}$ is non-unique and has a non-trivial
distribution. In other words, particle trajectories branch.

The same classification of the flows was obtained by Le Jan and
Raimond \cite{lera99} using Wiener chaos expansion without
explicit reference to the $\kappa$ limit. In contrast, our
primary motivation is to study the limit of physical regularizations.

The following result answers the question Q1 and also
points out that there are three different regimes if both the $\kappa$-
and the $\eps$-limits are considered.

\begin{theorem} 
  \label{th:1}
  In the strongly compressible regime when $\mathcal{P}\geq d/\xi^2$,
  there exists a two-parameter family of random maps
  $\{\flow^{\omega}_{t,s}({x})\}$, such that for all smooth test
  functions $\psi$ and for all $(s,t,{x})$, $s<t$,
  \begin{equation}
    \label{eq:c1}
    \mathbf{E} \left(S_{t,s}^{\omega,\kappa} \psi({x})-
    \psi(\flow^{\omega}_{t,s}({x}))\right)^2\to0,
  \end{equation}
as $\kappa\to0$,  and
\begin{equation}
  \label{eq:c2}
  \mathbf{E}  \left(\psi(\flow^{\omega,\eps}_{t,s}({x}))-
    \psi(\flow^{\omega}_{t,s}({x}))\right)^2\to0,
\end{equation}
as $\eps\to0$. Moreover, the limiting flow
$\{\flow^{\omega}_{t,s}({x})\}$ coalesces in the sense that for
almost all $(t,{x},{y})$, ${x}\not={y}$, we can
define a time $\tau$ such that $-\infty<\tau<t$ a.s. and
\begin{equation}
  \label{eq:coales}
  \flow_{t,s}^{\omega}({x}) = \flow_{t,s}^{\omega}({y})\quad \hbox{for }
  \ s\le\tau.
\end{equation}

In the weakly compressible regime when $\mathcal{P}\leq
(d+\xi-2)/2\xi$, there exists a random family of generalized flows
$g_\omega({x},t|d{y},s)$, such that for all test function $\psi$,
\begin{equation}
  \label{eq:Sgw}
  S^{\omega}_{t,s}\psi({x}) = \int_{\R^d} \psi({y})
  g_\omega({x},t| d{y}, s),
\end{equation}
satisfies
 \begin{equation}
    \label{eq:c3}
    \mathbf{E}\left(
      S_{t,s}^{\omega,\kappa} \psi({x})-
      S^{\omega}_{t,s}\psi({x})\right)^2\to0,
  \end{equation}
  as $\kappa\to0$ for all $(s,t,{x})$, $s<t$, and
 \begin{equation}
    \label{eq:c4}
    \mathbf{E}
    \Bigl(\int_{\R^d}
      \eta({x})\left(\psi(\flow^{\omega,\eps}_{t,s}({x}))-
      S^\omega_{t,s} \psi({x})\right)d{x}\Bigr)^2\to0,
  \end{equation}
  as $\eps\to0$ for all $(s,t)$, $s<t$, and for all test functions
  $\eta$.  Moreover, $g_\omega({x},t|d{y},s)$ is
  non-degenerate in the sense that
  \begin{equation}
    \label{eq:nondeg}
    S^\omega_{t,s} \psi^2({x})-
    \left(S^\omega_{t,s} \psi({x})\right)^2>0 \quad \hbox{a.s.}
  \end{equation}
  
  In the intermediate regime when $(d+\xi-2)/2\xi<
  \mathcal{P}<d/\xi^2$, there exists a random family of generalized
  flows $g_\omega({x},t|d{y},s)$, such that for all test function
  $\psi$ and for all $(s,t,{x})$, $s<t$,
  \begin{equation}
    \label{eq:c5}
    \mathbf{E}\left(S^{\omega,\kappa}_{t,s}\psi({x})-
    S^\omega_{s,t}\psi({x})\right)^2\to0
  \end{equation}
  as $\kappa\to0$. In the $\eps$-limit, the flows
  $\flow^{\omega,\eps}_{t,s}({x})$ converges in the sense of
  distributions, i.e. there exists a family of probability densities
  \begin{equation}
    \{G_n(x_1,\ldots,x_n,t|y_1,\ldots,y_n,s) dy_1\cdots dy_n \},
  \end{equation}
  $n=1,2,\ldots$, such that
\begin{equation}
  \label{eq:c6}
  \begin{array}{r}
    \d \mathbf{E} \psi(\flow^{\omega,\eps}_{t,s}({x_1}),
    \cdots,\flow^{\omega,\eps}_{t,s}({x_n}))\to
    \int\limits_{\R^d\times\cdots\times\R^d}\!\!\!\!
    \psi(y_1, \cdots, y_n)\
    \\[6pt]
    \d\times G_n(x_1,\cdots,x_n,t|y_1,\ldots,y_n,s) 
    dy_1 \cdots dy_n,
  \end{array}
\end{equation}
as $\eps\to0$ for any continuous function $\psi$ with compact support.
 Furthermore, the $\eps-$limit coalesces in the sense that
\begin{equation}
  \label{eq:c62}
  \begin{array}{r}
    \d G_2(x_1,x_2,t|y_1, y_2,s)=
    \tilde{G}_2(x_1,x_2,t|y_1, y_2,s)\qquad\\[6pt]
    \d \qquad+ A(y_1,x_1,x_2,t,s)\delta(y_1 - y_2),
  \end{array}
\end{equation}
with $A> 0$ when $t > s$. Here $\tilde{G}_2$ is the absolutely
continuous part of $G_2$ with respect to the Lebesgue measure.
Similar statements hold for the other $G_n$'s.  In particular, the
$\{G_n\}$'s differ from the moments of the $\kappa$-limit $g_\omega$
defined in \ref{eq:c5}.
\end{theorem}

Rephrasing the content of this result, we have strong convergence to a
family of flow maps in the strongly compressible regime for both the
$\kappa$-limit and the $\eps$-limit. In the weakly compressible
regime, we have strong convergence to a family of generalized flows
for the $\kappa$-limit, but weak convergence to the same limit for the
$\eps$-regularization. In fact, using the terminology of Young
measures \cite{tar83}, the limiting generalized flow
$\{g_\omega({x},t|d{y},s)\}$ is nothing but the Young measure
for the sequence of flow maps
$\{\varphi^{\omega,\eps}_{s,t}({x})\}$.  Finally, in contrast to
what is observed in the other two regimes, the $\eps$-limit and
$\kappa$-limit are not the same in the intermediate regime. As we will
see below, the structure functions of the passive scalar field
scale differently in the two limits.

 From Theorem~\ref{th:1}, it is natural to define the solution of
the transport equation in \ref{eq:scallimitLip1} for the initial
condition $\theta_\omega({x},s)=\theta_0({x})$ as
\begin{equation}
  \label{eq:stranps1}
  \theta_\omega({x},t) = S^\omega_{t,s} \theta_0({x})= 
  \int_{\R^d} \theta_0({y}) g_\omega({x},t|d{y},s),
\end{equation}
for the weakly compressible and the intermediate regimes in the
$\kappa$-limit (non-degenerate cases), and as
\begin{equation}
  \label{eq:stranps2}
  \theta_\omega({x},t) = \theta_0(\flow^{\omega}_{t,s}({x})),
\end{equation}
for the strongly compressible regime. In the intermediate regime in
the $\eps$-limit, it makes sense to look at the limiting moments of
$\theta^\eps_\omega({x},t)$ since we have as $\eps\to0$
\begin{equation}
  \label{eq:momentstheta}
  \begin{array}{r}
    \d \mathbf{E}(\theta^\eps_\omega({x}_1,t)\cdots 
    \theta^\eps_\omega({x}_n,t))\to\!\!\!\!
    \int\limits_{\R^d\times\cdots\times\R^d}\!\!\!\!
    \theta_0(y_1)\cdots\theta_0(y_n) \quad
    \\[6pt]
    \d\times G_n(x_1,\cdots,x_n,t|y_1,\ldots,y_n,s) 
    dy_1 \cdots dy_n.
  \end{array}
\end{equation}

It should be noted that when $g_\omega$ is non-degenerate, there
exists an anomalous dissipation mechanism for the scalar, whereas no
such anomalous dissipation is present in the coalescence cases
\cite{gave00}.  The presence of anomalous dissipation is the primary
reason why the transport equation in \ref{eq:scallimitLip1} has a
statistical steady state (invariant measure) if it is appropriately
forced, as we will show later.

Details of the proof of Theorem~\ref{th:1} are given in
\cite{evephysica}.  Crucial to the proof is the study of $P(\rho|r,
s)$ defined through $\eps$-regularization as
\begin{equation}
  \label{eq:Pdef}
    \int_{0}^\infty\eta(r) P(\rho|r,s-t)dr= \lim_{\eps\to0}
  \mathbf{E}\, \eta(|\flow^{\omega,\eps}_{t,s}({y})
      -\flow^{\omega,\eps}_{t,s}({z})|),
\end{equation}
where $\eta$ is a test function, and similarly through
$\kappa$-regularization.  Here $\rho=|{y}-{z}|$ and $s<t$. $P(\rho|r,
s)$ can be thought of as the probability density that two particles
have distance $r$ at time $s<t$ if their final distance is $\rho$ at
time $t$.  For Kraichnan model, $P$ satisfies the backward equation
\begin{equation}
  \label{eq:Peq}
  -\diff{P}{s}{}=- 
      \diff{}{r}{} \left(b(r)P\right)+\diff{}{r}{2} \left(a(r)P\right),
\end{equation}
for the final condition $\lim_{s\to0-}P(\rho|r,s)=\delta(r-\rho)$, and
with $a(r)$, $b(r)$ such that
\begin{equation}
  \label{eq:etadef}
  \begin{array}{c}
    {\d a(r) = D(\mathcal{S}^2+\xi\mathcal{C}^2) r^{\xi}+ O(r^2/\ell_0^2),}
    \\[6pt]
    {\d b(r)= D((d-1+\xi)\mathcal{S}^2-\xi\mathcal{C}^2) r^{\xi-1}
      + O(r/\ell_0^2).}
  \end{array}
\end{equation}
 For $r\gg\ell_0$, $a(r)$ tends to $C_0$, $b(r)$ to $ C_0(d-1)/r$, and
the equation in \ref{eq:Peq} reduces to a diffusion equation with
constant coefficient.  The equation in \ref{eq:Peq} is singular at
$r=0$.  The proof of Theorem~\ref{th:1} is essentially reduced to the
study of this singular diffusion equation.  This is also the main step
for which the white-in-time nature of the velocity field is crucial.

\section*{Structure Functions}

We now study some consequences of Theorem~\ref{th:1} for the passive
scalar $\theta_\omega$ defined in \ref{eq:stranps1} or
\ref{eq:stranps2}.  We note that the scaling of the second-order
structure function is the same for the $\kappa$- and the $\eps$-limits
in the strongly and the weakly compressible cases \cite{rem2}, but it
differs in the intermediate regime as a result of the difference
between the limits in \ref{eq:c5} and \ref{eq:c6}. For simplicity of
presentation, we assume that $\theta_0$ is isotropic and Gaussian.
Denote ($n\in\N$)
\begin{equation}
  \label{eq:struct}
   S_{2n}(|{x}-{y}|,t)=
   \mathbf{E}(\theta_\omega({x},t)-\theta_\omega({y},t))^{2n},
\end{equation}
$$
   \hbox{or}\quad S_{2n}(|{x}-{y}|,t)=\lim_{\eps\to0}
   \mathbf{E}(\theta^\eps_\omega({x},t)-\theta^\eps_\omega({y},t))^{2n},
  \eqno{({\bf\arabic{equation}}')}
$$
in the intermediate regime in the $\eps$-limit. In the strongly
compressible case, we have for both the $\kappa$- and the
$\eps$-limits
\begin{equation}
  \label{eq:scal22}
  S_2(r,t)= O(r^{\zeta}),
\end{equation}
with
\begin{equation}
  \label{eq:zeta}
  \zeta = \frac{2-d-\xi+2\xi\mathcal{P}}{1+\xi\mathcal{P}}.
\end{equation}
In the weakly compressible case, we have for both the $\kappa$- and
the $\eps$-limits
\begin{equation}
  \label{eq:scal12}
  S_2(r,t)
  =O(r^{2-\xi}).
\end{equation}
In the intermediate regime, the limits differ, and the $\kappa$-limit
scales as in \ref{eq:scal12}, whereas the $\eps$-limit scales as in
\ref{eq:scal22}.  The equations in \ref{eq:scal22} and \ref{eq:scal12}
can be derived upon expressing $s_2$ in terms of $P$; the details are
given in Ref.~\cite{evephysica}.

It is interesting to discuss the higher order structure functions both
in the non-degenerate and in the coalescence cases in \ref{eq:scal22}
and \ref{eq:scal12} since their scalings highlight very different
behavior of the scalar. We consider first the coalescence cases
which are simpler. In these cases, because of the absence of
dissipative anomaly, all higher order structure functions can again be
expressed in terms of $P$, and it can be shown
\cite{gave00} that
\begin{equation}
  \label{eq:scal2}
  S_{2n}(r,t) = O(r^{\zeta}),
\end{equation}
with $\zeta$ given by \ref{eq:zeta} for all $n\ge2$.  In fact,
coalescence implies that the temperature field~$\theta_\omega$ tends
to become flat except possibly on a zero-measure set where it presents
shock-like discontinuities. Such a situation with two kinds of spatial
structures for $\theta_\omega$ is usually refered to as bi-fractal,
and, in simple cases, one may identify $\zeta$ with the codimension of
the set supporting the discontinuities of $\theta_\omega$
\cite{pafr85,jaf97}.

The non-degenerate cases are more complicated.  In these cases,
one expects that $\theta_\omega$ presents a spatial behavior much
richer than in the coalescence cases, with all kinds of scalings
present. This is the multi-fractal situation for which the higher
order structure functions behave as
\begin{equation}
  \label{eq:scal1}
  S_{2n}(r,t)=O(r^{\zeta_{2n}}),
\end{equation}
with $\zeta_{2n} < n(2-\xi)$ for $2n>2$.  The actual value of the
$\zeta_n$'s cannot be obtained by dimensional analysis, and one has to
resort to various sophisticated perturbation techniques (see
\cite{shsi93,chfakole95,gaku95,bale98}).  We will
consider again the scaling of the structure functions at statistical
steady state in the section on incomplete self-similarity.

\section*{One Force, One Solution  Principle for Temperature}

We now turn to question Q3 and consider the existence of a statistical
steady state for the transport equation with appropriate forcing. We
restrict attention to the non-degenerate cases which include the
weakly compressible regime and the intermediate regime in the
$\kappa$-limit. Indeed, in these regimes the non-degeneracy of
$g_\omega({x},t|d{y},s)$ as a probability distribution in ${y}$
implies dissipation of energy or, phrased differently, decay in memory
in the semi-group $S_{t,s}$ generated by $\{g_\omega\}$. We show that
the anomalous dissipation is strong enough in order that the forced
transport equation has a unique invariant measure for both the weakly
compressible regime and the intermediate regime in the $\kappa$-limit.
This result, however, depends on the finiteness of $\ell_0$. In limit
as $\ell_0\to\infty$ an invariant measure exists only for the weakly
compressible regime.

We will consider (compare with \ref{eq:scallimitLip1})
\begin{equation}
  \label{eq:transpforce}
  \diff{\theta}{t}{}+({u}({x},t) \cdot \nabla) \theta = b({x},t).
\end{equation}
where $b$ is a white-noise forcing such that 
\begin{equation}
  \label{eq:Bdef}
  \mathbf{E}\, b({x},t) b({y},s)=B(|{x}-{y}|)\delta(t-s).
\end{equation}
$B(r)$ is assumed to be smooth and rapidly decaying to zero for $r\gg L$;
$L$ will be referred to as the forcing scale.  The solution of
\ref{eq:transpforce} for the initial condition
$\theta_\omega({x},s)=\theta_0({x})$ is understood as
\begin{equation}
  \label{eq:soltranspforce}
  \theta_\omega({x},t) = S^\omega_{t,s} \theta_0({x}) 
  + \int_{s}^{t} S^\omega_{t,\tau} b({x},\tau)d\tau.
\end{equation}
Define the product probability space
$(\Omega_u\times\Omega_b,\mathcal{F}_u\times\mathcal{F}_b,
\mathcal{P}_u\times\mathcal{P}_b)$, and the shift operator $T_\tau
\omega(t)=\omega(t+\tau)$, with $\omega=(\omega_u,\omega_b)$. We have

\begin{theorem}[One force--one solution I]
  \label{th:onefones}
  For $d>2$, in the weakly compressible regime and in the intermediate
  regime in the $\kappa$-limit, for almost all $\omega$, there exists a
  unique solution of \ref{eq:transpforce} defined on
  $\R^d\times(-\infty,\infty)$. This solution can be expressed as
\begin{equation}
  \label{eq:invmeas}
    \theta^\star_\omega({x},t) = 
    \int_{-\infty}^{t} S^\omega_{t,s} b({x},s)ds.
\end{equation}
 Furthermore the map $\omega\to\theta^\star_\omega$ satisfies the invariance
property
\begin{equation}
  \label{eq:inv}
  \theta^\star_{T_\tau \omega}({x},t) = 
  \theta^\star_{\omega}({x},t+\tau).
\end{equation}

\end{theorem}
Theorem~\ref{th:onefones} is the ``one force, one solution'' principle
articulated in \cite{ekhma99}. Because of the invariance property
\ref{eq:inv}, the map in \ref{eq:invmeas} leads to a natural invariant
measure. As a consequence we have

\begin{corollary}
  \label{th:invm}
  For $d>2$, in the weakly compressible regime and in the intermediate
  regime in the $\kappa$-limit, there exists a unique invariant measure
  on $L^2_\loc(\R^d\times \Omega)$ for the dynamics defined by
  \ref{eq:transpforce}.
\end{corollary}
The connection between the map \ref{eq:invmeas} and the invariant
measure, together with uniqueness, is explained in \cite{ekhma99}. The
restriction on the dimensionality in Theorem~\ref{th:onefones} arises
because the velocity field has finite correlation length~$\ell_0$:
Theorem~\ref{th:onefones} is changed into Theorem~\ref{th:onefonesv}
below in the limit as $\ell_0\to\infty$ which can be considered after
appropriate redefinition of the velocity field.

We sketch the proof of Theorem~\ref{th:onefones}. Basically, it
amounts to verifying that the dissipation in the system is strong
enough in the sense that
\begin{equation}
  \label{eq:proffth3s1}
  \mathbf{E} \Bigl(\int_{T_1}^{T_2} \int_{\R^d} S^\omega_{t,t+s}b({x},s) 
     ds \Bigr)^2 \to 0,
\end{equation}
as $T_1$, $T_2\to-\infty$ for fixed ${x}$ and $t$.
The average in \ref{eq:proffth3s1} is given
explicitly by
\begin{equation}
  \label{eq:proffth3s2}
  \int_{T_1}^{T_2} \int_0^\infty B(\rho) 
      P(0|\rho,s) drds,
\end{equation}
where $P$ satisfies \ref{eq:Peq}.  The convergence of the integral in
\ref{eq:proffth3s1} depends on the rate of decay in $|s|$ of
$P(0|\rho,s)$. The latter can be estimated by studying the equation in
\ref{eq:Peq} \cite{evephysica}, which yields $P(0|\rho,s)\sim C
\rho^{\alpha} |s|^{-d/2}$ with $\alpha =
(d-1-\xi(\xi+1)\mathcal{P})/(1+\xi\mathcal{P})$ for $|s|$ large and
$\rho\ll\ell_0$.  Hence, the integral in $s$ in \ref{eq:proffth3s2}
tends to zero as $T_1$, $T_2\to-\infty$ if $d>2$. It follows that the
invariant measure in \ref{eq:invmeas} exists provided $d>2$.

We now ask what happens if we let $\ell_0\to\infty$ in order to
emphasizes the effect of the inertial range of the velocity? This
question, however, has to be considered carefully because the velocity
field with the covariance in~\ref{eq:covdef} diverges as
$\ell_0\to\infty$.  The right way to proceed is to consider an
alternative velocity ${v}$, taken to be Gaussian, white-in-time, but
\textit{non-homogeneous}, with covariance
\begin{equation}
  \label{eq:covv}
  \begin{array}{l}
    {\d\mathbf{E}\, v_\alpha({x},t) v_\beta({y},s)}\\
    {\d=(c_{\alpha\beta}({x-a})+c_{\alpha\beta}({a-y})
      -c_{\alpha\beta}({x}-{y}))\delta(t-s).}
  \end{array}
\end{equation}
 For finite $\ell_0$, one has ${v}({x},t) = {u}({x},t)-{u}({a},t)$,
where ${a}$ is arbitrary but fixed. However, ${v}$ makes sense in the
limit as $\ell_0 \rightarrow \infty$.  Denote by
$\vartheta_\omega({x},t)$ the temperature field advected by ${v}$,
i.e. the solution of the transport equation \ref{eq:transpforce} with
${u}$ replaced by ${v}$:
\begin{equation}
  \label{eq:transpforcev}
  \diff{\vartheta}{t}{}+({v}({x},t) \cdot \nabla) \vartheta = b({x},t).
\end{equation}
Restricting to zero initial condition, it follows from the homogeneity
of the forcing that the single-time moments of $\theta_\omega$ and
$\vartheta_\omega$ coincide for finite $\ell_0$, but in contrast to
$\theta_\omega$, $\vartheta_\omega$ makes sense as $\ell_0\to\infty$.
Thus, $\vartheta_\omega$ is a natural process to study the limit as
$\ell_0\to\infty$, and we have
\begin{theorem}[One force--one solution II]
\label{th:onefonesv}
In the limit as $\ell_0\to\infty$ in the weakly compressible regime,
for almost all $\omega$, there exists a unique solution of
\ref{eq:transpforcev} defined on $\R^d\times(-\infty,\infty)$. This
solution can be expressed as
\begin{equation}
  \label{eq:invmeasv}
    \vartheta^\star_\omega({x},t) = 
    \int_{-\infty}^{t} S^\omega_{t,s} b({x},s)ds,
\end{equation}
where $S^\omega_{s,t}$ is the semi-group for the generalized flow
associated with the velocity defined in \ref{eq:covv} in the limit as
$\ell_0\to\infty$.  Furthermore the map
$\omega\to\vartheta^\star_\omega$ satisfies the invariance property
\begin{equation}
  \label{eq:invv}
  \vartheta^\star_{T_\tau \omega}({x},t) = 
  \vartheta^\star_{\omega}({x},t+\tau).
\end{equation}

\end{theorem}
As a direct result we also have
\begin{corollary}
  \label{th:invmv}
  In the limit as $\ell_0\to\infty$, in the weakly compressible regime
  there exists a unique invariant measure on $L^2_\loc(\R^d\times
  \Omega)$ for the dynamics defined by \ref{eq:transpforcev}.
\end{corollary}
Notice that, as $\ell_0\to\infty$, the anomalous dissipation is not
strong enough in the intermediate regime in the $\kappa$-limit, for
which no statistical steady state with finite energy exists.

The proof of Theorem~\ref{th:onefonesv} proceeds as the one for
Theorem~\ref{th:onefones}, but the estimate for $P$ in
\ref{eq:proffth3s2} changes as $P(0|\rho,s)\sim C \rho^{\alpha}
|s|^{-(\alpha+1)/(2-\xi)}$ with $\alpha =
(d-1-\xi(\xi+1)\mathcal{P})/(1+\xi\mathcal{P})$ for $|s|$ large and
$\rho\ll\ell_0$.  It follows that the integral in $s$ in
\ref{eq:proffth3s2} converges as $T_1$, $T_2\to-\infty$ in the weakly
compressible regime only.

\section*{One Force, One Solution  Principle for the Temperature Difference}

Since no anomalous dissipation is present in the coalescence
cases, i.e the stron\-gly compressible regime and the intermediate
regime in the $\eps$-limit, no invariant measure for the temperature
field exists in these regimes. It makes sense, however, to ask about
the existence of an invariant measure for the temperature difference,
i.e. to consider
\begin{equation}
  \label{eq:invmeasd}
    \delta\theta_\omega({x},{y},t) = 
    \int_{T}^{t} S^\omega_{t,s} (b({x},s)-b({y},s))ds,
\end{equation}
in the limit as $T\to-\infty$.  When $\theta^\star_\omega$ exists, one
has
\begin{equation}
  \label{eq:deltathetas}
  \delta\theta^\star_\omega({x},{y},t)=
  \lim_{T\to-\infty}\delta\theta_\omega({x},{y},t)=
  \theta^\star_\omega({x},t)-\theta^\star_\omega({y},t),
\end{equation}
but it is conceivable that $\delta\theta_\omega^\star$ exists in the
coalescence cases even though $\theta^\star_\omega$ is not defined.
The reason is that coalescence of the generalized flow implies that
the temperature field flattens with time, which is a dissipation
mechanism as far as the temperature difference is concerned. Of
course, this effect has to overcome the fluctuations produced by the
forcing, and the existence of an invariant measure such as
\ref{eq:invmeasd} will depend on how fast particles coalesce under the
flow, which happens only in the limit as $\ell_0\to\infty$ (i.e. for
the alternate velocity defined in~\ref{eq:covv}) as we show now.

 For finite $\ell_0$, if we were to consider two particles separated by
much more than the correlation length $\ell_0$, the dynamics of their
distance under the flow is governed by the equation in \ref{eq:Peq}
for $r\gg\ell_0$, i.e. by a diffusion equation with constant diffusion
coefficient on the scale of interest.  It follows that no tendency of
coalescence is observed before the distance becomes smaller than
$\ell_0$, which, as shown below, does not happen fast enough in order
to overcome the the fluctuations produced by the forcing. In other
words,

\begin{lemma}
  \label{th:noSSS}
  In the coalescence cases, for finite $\ell_0$, there is no
  invariant measure with finite energy for the temperature difference.
\end{lemma}

Consider now the limit as $\ell_0\to\infty$, and let
$\delta\vartheta_\omega({x},{y},t)=
\vartheta_\omega({x},t)-\vartheta_\omega({y},t)$ where
$\vartheta_\omega$ solves the equation in \ref{eq:transpforcev}.  The
temperature difference $\delta\vartheta_\omega$ satisfies the
transport equation
\begin{equation}
  \label{eq:transpforce2}
  \diff{\delta\vartheta}{t}{}
  +({v}({x},t) \cdot \nabla_x 
  +{v}({y},t) \cdot \nabla_y) \delta\vartheta
  = b({x},t)-b({y},t).
\end{equation}
We have

\begin{theorem}[One force--one solution III]
\label{th:onefones2}
In the limit as $\ell_0\to\infty$, for almost all $\omega$, in the
strongly and the weakly compressible regimes, as well as in the
intermediate regime if the flow is non-degenerate, there exists a
unique solution of \ref{eq:transpforce2} defined on
$\R^d\times(-\infty,\infty)$. This solution can be expressed as
\begin{equation}
  \label{eq:invmeas2}
    \delta\vartheta^\star_\omega({x},{y},t) = 
    \int_{-\infty}^{t} S^\omega_{t,s} (b({x},s)-b({y},s))ds,
\end{equation}
where $S_{s,t}$ is the semi-group for the generalized flow associated
with the velocity defined in \ref{eq:covv} in the limit as
$\ell_0\to\infty$.  Furthermore the map
$\omega\to\delta\vartheta^\star_\omega$ satisfies the invariance
property
\begin{equation}
  \label{eq:inv2}
  \delta\vartheta^\star_{T_\tau \omega}({x},{y},t) 
  = \delta\vartheta^\star_{\omega}({x},{y},t+\tau).
\end{equation}
\end{theorem}
An immediate consequence of this theorem is 

\begin{corollary}
  \label{th:invm2}
  In the limit as $\ell_0\to\infty$, in the strongly and the weakly
  compressible regimes, as well as in the intermediate regime if the
  flow is non-degenerate, there exists a unique invariant measure on
  $L^2_\loc(\R^d\times \Omega)$ for the dynamics defined by
  \ref{eq:transpforce2}.
\end{corollary}
The proof of Theorem~\ref{th:onefones2} proceeds similarly as the
proof of Theorem~\ref{th:onefones}. In the non-degenerate cases,
one study the convergence of (compare \ref{eq:proffth3s1})
\begin{equation}
  \label{eq:proffth3vds1}
  \mathbf{E} \Bigl(\int_{T_1}^{T_2} \int_{\R^d} S^\omega_{t,t+s} 
  (b({x},s)-b({y},s)) ds \Bigr)^2 \to 0,
\end{equation}
as $T_1$, $T_2\to-\infty$ for fixed ${x}$ and $t$.  The average in
\ref{eq:proffth3vds1} can be expressed in terms of $P$, and it can be
shown \cite{evephysica} that the expression in (\ref{eq:proffth3vds1})
converges as $T_1$, $T_2\to-\infty$ in the non-degenerate cases.
In the strongly compressible regimes, because of the existence of a
flow of maps, \ref{eq:proffth3vds1} is replaced by
\begin{equation}
  \label{eq:proffth3s12}
  \mathbf{E} \Bigl(\int_{T_1}^{T_2} ( b(\flow^{ \omega}_{t,s}({x}),s) 
    -b(\flow^{ \omega}_{t,s}({y}),s)) ds \Bigr)^2.
\end{equation}
This average can again be expressed in terms of $P$, and it can be
shown that the convergence of the time integral in
\ref{eq:proffth3s12} depends on the rate at $P$ looses mass at $r=0+$
(i.e.  the rate at which particles coalesce). The analysis of the
equation in \ref{eq:Peq} shows that the process is fast enough in
order that the integral over $s$ in \ref{eq:proffth3s12} tends to zero
as $T_1$, $T_2\to-\infty$ in the strongly compressible regime. In
contrast, the equivalent of \ref{eq:proffth3s12} in the intermediate
regime in the $\eps$-limit can be shown to diverge as $T_1$,
$T_2\to-\infty$.

It can be shown that the invariant measure has finite correlation
functions of all order, even though these results do not by themselves
imply uniqueness of stationary solutions to the $n$-point
 Fokker-Planck equation. The task of studying the passive scalar is now
changed to the study of the short distance behavior of these
correlation functions.

\section*{Incomplete self-similarity}

We finally turn to question Q4 and consider the scaling of the
structure functions based on the invariant measure $\delta\vartheta^\star$
defined in \ref{eq:invmeas2}. Denote
\begin{equation}
  \label{eq:strucfct}
  \mathcal{S}_{n} (|{x}-{y}|) 
  = \mathbf{E} \, |\delta\vartheta_\omega^\star({x},{y},t)|^{n}.
\end{equation}
The dimensional parameters are $B_0=B(0)$
($[$temperature$]^2$\-$[$time$]^{-1}$),\\ $D$
($[$length$]^{2-\xi}$\-$[$time$]^{-1}$), $L$ ($[$length$]$). It follows
that
\begin{equation}
  \label{eq:anscaling}  
  \mathcal{S}_n(r)= \left(\frac{B_0r^{2-\xi}}{D}\right)^{n/2} 
  f_n\left(\frac{r}{L}\right),
\end{equation}
where the $f_n$'s are dimensionless functions which cannot be obtained
by dimensional arguments.  For instance, the scalings in
\ref{eq:scal2}, \ref{eq:scal1} correspond to different $f_n$. It is
however obvious from the equation \ref{eq:anscaling} that, provided
the limit exists and is non-zero
\begin{equation}
  \label{eq:regscaling}
  \lim_{L\to\infty} \mathcal{S}_n(r)
  = C_{n} \left(\frac{B_0r^{2-\xi}}{D}\right)^{n/2}= O(r^{n(2-\xi)/2}).
\end{equation}
where $C_n= \lim_{r\to\infty}f_n(r/L)$ are numerical constants. The
scaling in \ref{eq:regscaling} is usually referred to as the normal
scaling since, consistent with Kolmogorov's picture, it is independent
of the forcing or the dissipation scales.  In contrast, anomalous
scaling is a statement that the structure functions diverge in the
limit of infinite forcing scale, $L\to\infty$.  In the spirit of
Barenblatt-Chorin \cite{bach97,bach98}, we may say that normal
scaling holds in case of complete self-similarity, whereas anomalous
scaling is equivalent to incomplete self-similarity.

It is interesting to discuss the existence or non-existence of the
limit in \ref{eq:regscaling} for both the coalescence and the
non-degenerate cases. When the flow coalesces, because of the
existence of a flow of maps and the absence of dissipative anomaly,
the $\mathcal{S}_{2n}$'s of even order $2n\ge 2$ can be computed
exactly \cite{gave00}. It gives $\mathcal{S}_{2n}(r)=\infty$ for $n\ge
\zeta/(2-\xi)$, whereas
\begin{equation}
  \label{eq:anscalingc}  
  \mathcal{S}_{2n}(r)= O(r^\zeta)\qquad 
  \hbox{for } \ n < \frac{\zeta}{2-\xi},
\end{equation}
where $\zeta$ is given in~\ref{eq:zeta}. Thus, for $n < \zeta/(2-\xi)$,
\begin{equation}
  \label{eq:fnc}
  f_{2n}(r) = O\left((r/L)^{\zeta-n(2-\xi)}\right).
\end{equation}
It follows that $f_{2n}$ and, hence, $\mathcal{S}_{2n}$ tend to zero
as $L\to\infty$ for $2\le n < \zeta/(2-\xi)$, whereas they are
infinite for all $L$ for $n\ge \zeta/(2-\xi)$. In fact, in the
coalescence case, it can be shown \cite{gave00} that on scales much
larger than the forcing scale $L$, the structure functions of order
$n< \zeta/(2-\xi)$ behave as
\begin{equation}
  \label{eq:anscalingc2}  
  \mathcal{S}_{2n}(r) \sim C_{2n} r^{n(2-\xi)} \qquad 
  \hbox{as } \ r/L\to\infty.
\end{equation}
Thus in the coalescence case, it is more natural to consider the
limit as $L\to0$ of the structure functions, for which the expression
in \ref{eq:anscalingc2} shows the absence of intermittency
corrections.

In the non-degenerate case, one has 
\begin{equation}
  \label{eq:S2d}
  \mathcal{S}_2(r) = O(r^{2-\xi}), 
\end{equation}
while perturbation analysis gives for the higher order structure
functions \cite{shsi93,chfakole95,gaku95,bale98}
\begin{equation}
  \label{eq:S2nd}
  \mathcal{S}_{2n}(r) = O(r^{\zeta_{2n}}), 
\end{equation}
with $\zeta_{2n} < n(2-\xi)$ for $2n>2$. It follows that $f_{2}(r) =
O(1)$, while
\begin{equation}
  \label{eq:fnd}
  f_{2n}(r) = O\left((r/L)^{\zeta_n-n(2-\xi)}\right), \qquad 2n>2.
\end{equation}
In other words, as $L\to\infty$, $\mathcal{S}_2$ has a limit which
exhibits normal scaling, whereas the $\mathcal{S}_{2n}$'s, $2n>2$,
diverge. This may be closely related to the argument in
\cite{bach97,bach98} that, in appropriate limits, intermittency
corrections may disappear and higher than fourth order structure
functions may not exist. We note, however, that Barenblatt and Chorin
were discussing the case of infinite Reynolds number (here infinite
Peclet number, $\kappa\to0$) at finite $L$, whereas we require
$L\to\infty$.

~

We thank many people for helpful discussions, including G.
Barenblatt, E. Balkovsky, M. Chertkov, A. Chorin, U. Frisch, J.
Goodman, A. Majda, and S. R. S. Varadhan.  We are particularly
grateful to K.  Gaw\c edzki for pointing out an error in the first
draft of this paper. W. E is partially supported by a Presidential
Faculty Fellowship from NSF.  E. Vanden Eijnden is partially supported
by NSF Grant DMS-9510356.


\begin{thebibliography}{99}
\bibitem{rem1} As long as ${u}$ is continuous, Peano's theorem tells us
  that solutions to \ref{eq:deter} do exist.
\bibitem{rem2} For the weakly compressible case, because the
  $\eps$-limit is a weak limit one has to consider the structure
  functions of the mollified temperature field $\psi_\eps \star
  \theta^\eps$ in order to obtain the limiting scaling as $\eps\to0$.
\end{thebibliography}
\end{document}